\documentclass[conference,compsoc]{IEEEtran}
\IEEEoverridecommandlockouts
\usepackage{cite}
\usepackage{amsmath,amssymb,amsfonts}
\usepackage{algorithmic}
\usepackage{graphicx}
\usepackage{hyperref}
\usepackage{textcomp}
\usepackage{xcolor}
\usepackage{booktabs}
\usepackage[capitalise]{cleveref}
\def\BibTeX{{\rm B\kern-.05em{\sc i\kern-.025em b}\kern-.08em
    T\kern-.1667em\lower.7ex\hbox{E}\kern-.125emX}}
\begin{document}

\title{What Determines the Price of NFTs?}

\author{\IEEEauthorblockN{Vivian Ziemke, Benjamin Estermann, Roger Wattenhofer and Ye Wang$^*$}
\IEEEauthorblockA{\{ziemkev,	besterma,wattenhofer\}@ethz.ch, wangye@um.edu.mo\\
ETH Zurich, University of Macau}
\IEEEauthorblockA{$^*$Corresponding author.}
}

\maketitle

\begin{abstract}
In the evolving landscape of digital art, Non-Fungible Tokens (NFTs) have emerged as a groundbreaking platform, bridging the realms of art and technology.
NFTs serve as the foundational framework that has revolutionized the market for digital art, enabling artists to showcase and monetize their creations in unprecedented ways.
NFTs combine metadata stored on the blockchain with off-chain data, such as images, to create a novel form of digital ownership.
It is not fully understood how these factors come together to determine NFT prices.
In this study, we analyze both on-chain and off-chain data of NFT collections trading on OpenSea to understand what influences NFT pricing.
Our results show that while text and image data of the NFTs can be used to explain price variations within collections, the extracted features do not generalize to new, unseen collections.
Furthermore, we find that an NFT collection's trading volume often relates to its online presence, like social media followers and website traffic.


\end{abstract}


\section{Introduction}

In recent years, Non-Fungible Tokens (NFTs) have transformed the realm of digital art, offering artists an unparalleled opportunity to exhibit and profit from their creations. NFTs introduce an innovative framework that has ushered in a new era for digital art by empowering artists to establish ownership and authenticity in an increasingly digital world. In 2022, the NFT market recorded a remarkable trading volume of 24.7 billion USD~\cite{TradingVolumeNFT}, emphasizing the significant influence of NFTs on the creative landscape.

This transformation in the art world has prompted a pressing question: What factors drive NFT prices in this dynamic and ever-evolving market? The discourse surrounding NFT valuation is multifaceted, with some contending that prices are primarily speculative, detached from the intrinsic worth of the underlying artwork, while others assert that artistic merit significantly influences market values.

While the blockchain serves as the ledger for recording NFT ownership and transactional data, our focus in this study is on the artists, their creations and other off-chain data, rather than on the blockchain itself. NFTs, as the pioneers in creating a market for digital art, have redefined the relationship between artists and their audience. This paper delves into the heart of these dynamics, aiming to unravel the determinants of NFT prices. To achieve this, we address the following research question:

\textbf{RQ:} Which factors determine the price and trading volume of NFTs?

We compile two data sets that include both on-chain and off-chain data for NFT transactions carried out on OpenSea markets from 2017 until January 2021. We also enhance our analysis by utilizing market data provided by Nadini et al.~\cite{nadini2021mapping}. Our initial approach is to develop machine learning models to predict NFT prices based on text descriptions and associated images. Subsequently, we expand the model's feature set to include metrics such as social media traction, keyword search frequencies, and website visits. This allows us to determine the influence of these off-chain attributes on NFT pricing.

Our study sheds light on multiple facets of NFT pricing mechanisms. Using a machine learning model utilizing a Bag-of-Words approach, we demonstrate the significant impact of textual descriptions on NFT valuation, identifying specific keywords in collections that have a substantial effect on pricing. While image characteristics are correlated with NFT prices, predicting prices for NFTs within large collections based solely on image data proves difficult. These findings underscore the presence of distinct, non-visual determinants that influence prices in specific collections. Additionally, off-chain data-particularly metrics such as Twitter followers and recent website traffic-show a robust association with trading volumes. 

\section{Background \& Related Work}

Blockchain technology, ever since its inception, has played a disruptive role in the digital arena. Initially conceptualized as an underlying structure for Bitcoin, the technology has far outgrown its primary use-case, heralding a new era of decentralized applications \cite{nakamoto2008bitcoin}. At its core, blockchain is a decentralized ledger, immutable in nature, ensuring that once data is written onto it, it becomes nearly impossible to change without a consensus. This decentralized, trustless architecture ensures transparency and security, paving the way for novel digital assets without the need for intermediaries \cite{tapscott2016blockchain}.

One of the most revolutionary assets enabled by the capabilities of blockchain technology is the Non-Fungible Token (NFT). Unlike traditional cryptocurrencies like Bitcoin or Ethereum, which are fungible and where every token is identical to every other, NFTs are distinct. Each NFT is unique and indivisible, representing a specific item or piece of content on the blockchain \cite{kugler2021non}. This uniqueness is what makes NFTs particularly suitable for digital art and collectibles. Artists can create tokens of their work to provide buyers with a digital certificate of authenticity, which guarantees ownership and rarity of the artwork piece. NFTs have enabled digital art to assert value, rarity, and provenance, which were previously challenges in the digital art world \cite{dowling2022non, dowling2022fertile}.

The NFT marketplace diverges significantly from traditional art markets, presenting unique challenges and dynamics. Extensive research has explored the high volatility in NFT pricing, identifying a myriad of contributing factors~\cite{dowling2022non, luo2022understanding, kapoor2022tweetboost, schnoering2022constructing, nadini2021mapping, white2022characterizing}. Notably, the overarching cryptocurrency market trends and specific NFT market activities have been found to exert a substantial influence on NFT valuations~\cite{dowling2022non, schnoering2022constructing}. Additionally, the impact of social media presence, particularly activity on platforms like Twitter, has been robustly correlated with fluctuations in NFT market prices~\cite{kapoor2022tweetboost}. These external forces, often unrelated to the inherent artistic value of the NFT, induce notable short-term price variability. This highlights the NFT market's fluid, multifaceted, and at times capricious nature, which adds a layer of complexity when valuing digital art~\cite{luo2022understanding, nadini2021mapping}.

In the burgeoning domain of machine learning applications for NFT market analysis, existing studies have begun to tap into the vast array of data available, but certain limitations persist. For instance, Kapoor et al. honed in on the correlation between Twitter activities and NFT valuations, yet their research lacks generalizability to previously unseen NFT collections, constraining the applicability and robustness of their models~\cite{kapoor2022tweetboost}. Costa et al. utilized deep learning techniques to predict NFT prices using image and text-based data, but their scope is circumscribed by a three-month data window, undermining the temporal validity of their conclusions~\cite{costa2023show}. Our research addresses these limitations by leveraging a dataset that spans from 2017 to January 2021, thereby providing a more comprehensive and temporally nuanced perspective on NFT market dynamics.

\section{Data Collection}
We analyze the influence of the artwork and off-chain data using two distinct datasets.
\subsection{Effect of artwork}
Initially, we capitalize on the data curated by \cite{nadini2021mapping} to construct a dataset comprising 10,000 NFT images, accompanying text descriptions, and corresponding prices denominated in ETH.

\subsubsection{Preprocessing}
\begin{figure}
    \centering
    \includegraphics[width=\columnwidth]{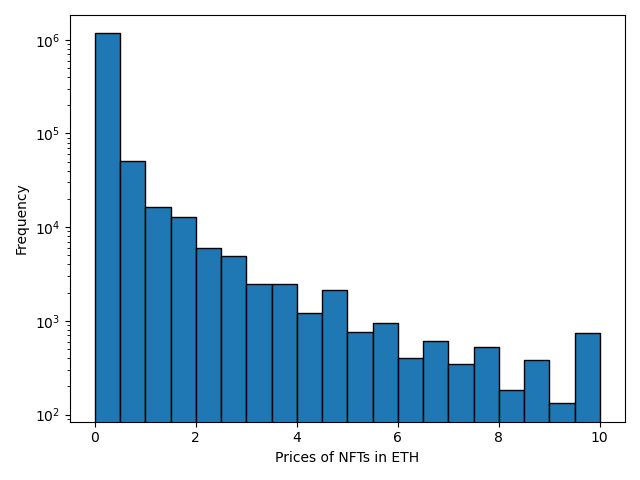}
    \caption{Distribution of NFT prices after outlier removal but before normalization, where the y-axis is in log scale.}
    \label{fig:price-dist-unfiltered}
\end{figure}

\begin{figure}
    \centering
    \includegraphics[width=\columnwidth]{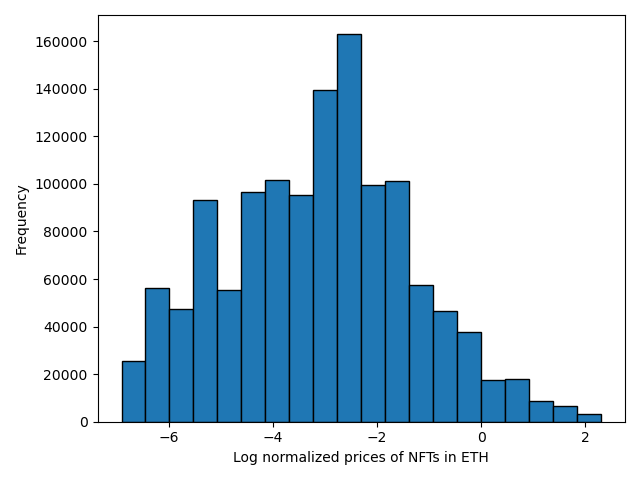}
    \caption{Distribution of NFT prices after log normalization.}
    \label{fig:price-dist-log}
\end{figure}

The data preprocessing entails a two-step procedure.
First, to ensure uniformity, we exclude transactions that didn't occur via the Ethereum blockchain, thus enabling a consistent comparison of prices in Ether coin currency.
Subsequently, we trim any transaction price outliers located more than 3 standard deviations from the mean price. 
Consequently, all remaining transactions are confined within the 0.001 to 10 ETH range.
Notably, visualizing the price distribution reveals a stark skewness, as depicted in \cref{fig:price-dist-unfiltered}.
To address this skewness and create a more balanced distribution, we perform a log transformation on the price data.
This transformation results in a distribution that approximates a bell curve, as shown in \cref{fig:price-dist-log}.

For the textual descriptions of NFTs, we utilize the sklearn TfidfVectorizer.
This tool converts these descriptions into Bag-of-Words representations. 
In simpler terms, it generates a list of all distinct words used in the descriptions and counts the frequency of each word's occurrence.
Each description is then transformed into a binary array, indicating whether each word is present or not.
Additionally, common English stop words, such as ''the,`` ''is,`` and ''and,`` are filtered out due to their lack of substantive meaning.


\subsection{Effect of off-chain data}
Our approach for building the collection dataset begins with scraping data from the OpenSea collection statistics webpage.
We focus on Ethereum collections, sort them by lifetime trading volume, and capture the top 1000 results \cite{OpenSeaCollectionRanking}.
Subsequently, we enhance this dataset by utilizing the OpenSea API to gather more detailed information, including creation dates, collection websites, URLs, and social media links. 
These social media links enable us to retrieve follower counts for each collection. 
Additionally, we extract category information from the OpenSea website and incorporate it into our dataset.
It is important to note that some collections lack social media accounts or a specified category, which we denote using negative values in the dataset and exclude from our analysis.

Alongside this, we compile historical data related to recent website traffic associated with NFT collections.
We leverage the Zylas Site Traffic API \cite{zylaapi} to obtain estimates of website visitor numbers and their nationalities over the past three months.

To create the historical trading timeseries dataset, we calculate monthly trading volumes from January 2018 to May 2023 for each collection in the collection dataset.
Given that Ethereum transaction data is available from multiple API providers, we utilize the OpenSea API Event Endpoint \cite{OpenSeaBlocks} and specify ''event\_type=Successful`` to capture successful transactions.
We compute Unix timestamps for the current and subsequent months, which delineate the transaction period.
Due to the API's constraint of returning a maximum of 20 results, we iterate through each month, adjusting the starting timestamp incrementally until fewer than 20 results are obtained or the month's end timestamp is reached.

To complement the historical trading data, we collect historical data related to NFT collections using keyword searches. To gauge general internet interest, we turn to Google Trends, which provides historical data on keyword searches conducted via the Google search engine. Our method involves configuring Selenium to set the region as global, the time span as the past five years, and inputting the relevant keywords.
Formulating optimal keywords from collection names poses a challenge, balancing the need for results against potential overlap with unrelated topics sharing the same name.
To this end, we devise a keyword generation process: we remove "official" from the collection name, eliminate punctuation and special characters, and query the keyword alone and in combination with "NFT." 
The latter step ensures that the search is NFT-oriented, although it might yield reduced data.
In cases where the "NFT" query yields no data, we revert to the collection name query alone. 
This approach seeks to maximize the relevance of search results to NFT collections while avoiding extraneous noise.

\section{Methods}
\subsection{Text Description Analysis}
    \label{subsec:methods-text}
    To assess text similarity both within and between collections, we calculate pairwise cosine similarities. 
    The computed pairwise cosine similarity within collections stands at 0.545, which is notably higher—about five times—than the cosine similarity value of 0.109 observed between different collections.
    This observation leads us to hypothesize that machine learning models should be adept at detecting these inherent similarities.
    
    We proceed by training and contrasting six diverse machine learning models: Linear Regression, Ridge Regression, Lasso Regression, and Decision Tree, all using default hyperparameter settings.
    Additionally, we introduce a baseline method that predicts the average price of the NFT collection it belongs to.

    We provide brief explanations of the machine learning models utilized:
    \begin{itemize}
        \item \textbf{Linear Regression: }A foundational model that establishes linear relationships between input features and the target variable, aiming to minimize the residual sum of squares.
        \item \textbf{Ridge Regression: }A variant of linear regression that introduces L2 regularization, aiding in preventing overfitting by penalizing large coefficient values.
        \item \textbf{Lasso Regression: }Similar to Ridge, Lasso incorporates regularization (L1) that not only prevents overfitting but also induces feature selection by pushing some coefficients to zero.
        \item \textbf{Decision Tree: }A non-linear model that forms a tree-like structure by recursively partitioning the feature space based on feature values, making predictions through majority voting in the leaf nodes.
    \end{itemize}

    These models, each with its distinct characteristics, enable us to comprehend how well they capture the relationships between the Bag-of-words representation and the NFT's prize, both in terms of in-distribution and out-of-distribution cases.

    Our analysis encompasses two distinct scenarios, where both datasets are divided into an 80\% training set and a 20\% test set.
    In the first case, the dataset is randomly partitioned into training and testing subsets. 
    We term this the \textbf{in-distribution case}.
    In contrast, the second case involves splitting the dataset so that collections present in the training set are entirely absent from the test set.
    In other words, the collections featured in the test set are entirely novel in the context of training.
    This configuration allows us to evaluate whether the acquired features generalize to unseen collections, and we term it the \textbf{out-of-distribution case}.

\subsection{Image Analysis}
    To explore how NFT images affect pricing, we made and compared different machine learning models that predict NFT prices using their images.
    We used 10,000 NFT images for this study.
    We used popular pre-trained Convolutional Neural Network (CNN) models: VGG16, ResNet50, InceptionV3, DenseNet121, EfficientNetB0, and Xception. 
    These models were trained on the Imagenet dataset \cite{imagenet} to recognize objects and classify images.
    We loaded these pre-trained CNN models but excluded their top layer.
    We kept their learned features unchanged and didn't modify their core structure.
    Then, we added a global average pooling layer, a dense layer with 128 neurons, and a single-neuron output layer to predict prices.
    We trained these custom layers using the Adam optimizer, aiming to reduce the mean squared error, which is a good measure for predicting prices. 
    We use the same two train-test splits described in \cref{subsec:methods-text}, where the second split allows us to see if the identified image features generalize to unseen collections.

    The decision to employ pre-trained CNN models stems from their capacity to extract intricate features from images. 
    They learned from a large and diverse dataset to recognize things like edges, textures, and parts of objects. 
    By using these pre-trained models and extending their feature extraction capabilities, we are able to use the wealth of image-relevant information to predict NFT values.


\subsection{Off-chain Data Analysis}
    In our off-chain data analysis, we focus on the total trading volume of NFT collections – the sum of transactions involving those collections. 
    This total volume reflects the collection's economic performance, as creator earnings from resales can be calculated by multiplying the volume with the creator fee.
    
    Before analysis, we remove outliers. 'Rarible' is more of a marketplace than a collection, distorting the dataset.
    'The 140 Collection by Twitter' is problematic due to the loose connection to of NFTs, causing inflated Twitter follower counts.
    A comparable challenge arises from 'The 140 Collection by Twitter,' where the connection between the NFT collection and the official Twitter account, under the handle "Twitter," remains tenuous. This disconnection leads to an unjustified inflation of the Twitter follower counts, warranting its exclusion.
    Furthermore, we compare website traffic visitor from the past three months to the total trading volumes of collections from their whole lifetime. 
    We calculate Pearson and Spearman correlation coefficients with corresponding p-values for all features.
    These coefficients help us understand how trading volume correlates with off-chain factors that might affect NFT prices.
    Pearson gauges linear relationships, while Spearman suits non-linear cases or rank-based data.
    These coefficients quantify how trading volume and off-chain factors relate, offering insights into factors impacting NFT prices.

    In addition to the monthly trading volume data, we have also amassed a dataset containing monthly Google Trends interest metrics. This augmentation of data enables us to conduct a more comprehensive analysis. Specifically, our investigation entails an exploration of cross-correlation lags.

    Cross-correlation lags, in this context, signify the time intervals between fluctuations in the trading volume and the corresponding shifts in Google Trends interest levels. In essence, these lags offer insights into the temporal relationship between two sets of data. A positive cross-correlation lag indicates that a surge in trading volume tends to be followed by an increase in Google Trends interest, while a negative lag implies a delayed rise in interest subsequent to trading volume spikes. Evaluating these cross-correlation lags enhances our understanding of how changes in trading volume and Google Trends interest align chronologically, thus shedding light on potential cause-and-effect dynamics or co-occurring trends. This analysis helps uncover potential dependencies between the popularity of NFT collections and broader online search behaviors.

\section{Results}
    
    \subsection{Effect of Artwork}
    We measure and compare the models with R2 scores. 
    The R2 score, also known as the coefficient of determination, is a statistical measure that gauges the proportion of variance in the dependent variable (in this case, NFT prices) that can be explained by the independent variables (such as text descriptions or image features) used in the model. 
    The R2 score ranges from 0 to 1, with higher values indicating that the model's predictions are more in line with the actual prices.
    An R2 score of less or equal to 0 implies that the model does not explain any variance, while an R2 score of 1 indicates that the model perfectly predicts the prices.
    Therefore, when we say an R2 score is 0.2, it means the model can account for around 20\% of the variability in prices using the provided information.
    
        \subsubsection{Text}
            Looking at the results of the in-distribution case, our baseline model achieves a R2 score of 0.266, meaning it can predict roughly one quarter of the variance in the prices from the average prices of each collection.

            \begin{table}[htbp]
              \centering
              \caption{R2 Scores by Model and Normalization Type  with standard train/test split of the BagOfWord ML models from NFT descriptions.}
              \begin{tabular}{cc}
                \toprule
                \textbf{Model} & \textbf{R2 Score} \\
                \midrule
                Linear Regression &  -0.475621 \\
                Ridge Regression & \textbf{0.539066} \\
                Lasso Regression &  -0.000046 \\
                Decision Tree Regression  & 0.337074 \\
                Baseline & 0.266 \\
                \bottomrule
              \end{tabular}
              
              \label{tab:r2-text-id}
            \end{table}
    
            \Cref{tab:r2-text-id} shows the R2 scores of the different Bag-of-Words models. The best performing model is the ridge regression model, outperforming our baseline significantly. 
            These results show that the model was able to identify some parts of the description which influence the price of an item within a collection the most.
            Most likely, the model also learned to identify to which collection an item belongs to, as this is also a strong indicator of the price.
            In \cref{tab:words-imprtance}, we display the most important words, as extracted from the ridge regression model.

            \begin{table}[ht]
                \centering
                \caption{Weighted words of the Bag-of-Words representation trained on ML models with collection info and average price}
                \begin{tabular}{ccc}
                \toprule
                \textbf{word} & \textbf{weight} & \textbf{collections} \\
                \midrule
                \ lilbun & 7.77   & Cryptokitties    \\
                \ dingtush & 6.887 &  Cryptokitties  \\
                \ sulkyki &  6.85 & Cryptokitties  \\
                 \ kingwuv & 6.64 & Cryptokitties  \\
                 \ gen & 6.41 & Cryptokitties, Avastar, etc.  \\
                 \ couple & 6.26 & Cryptokitties, Rarible, etc.  \\
                 \ sunnytoes & 6.05 & Cryptokitties  \\
                 \ blance & 5.82 & Cryptokitties \\
                 \ gen0 & 5.70 & Cryptokitties, Fydcards, Cryptomotors  \\
                \bottomrule
                \end{tabular}
            
                \label{tab:words-imprtance}
            \end{table}
    
            When looking at the results of the out-of-distribution case presented in \cref{tab:r2-text-ood}, it is evident that the text description features do not generalize to unseen collections. 
            This is display by consistently negative R2 scores.
            Therefore we conclude that most of the text descriptions can only be related to pricing within the context of the same collection.
            A further indicator of this hypothesis is that the pairwise cosine similarity within collections is 0.545, 5 times higher than the cosine similarity of 0.109 between collections.
            
            \begin{table}[htbp]
              \centering
              \caption{R2 Scores by Model and Normalization Type when the train/test split does not divide collections, showing that the model is not able to generalize to new unseen collections.}
              \begin{tabular}{cc}
                \toprule
                \textbf{Model} & \textbf{R2 Score} \\
                \midrule
                Linear Regression & -5.330842 \\
                Ridge Regression & -0.064147 \\
                Lasso Regression & -0.028932 \\
                Decision Tree Regression & -0.557529 \\
                Baseline & -0.191 \\
                \bottomrule
              \end{tabular}
              
              \label{tab:r2-text-ood}
            \end{table}

        \subsubsection{Images}
        Shifting our focus to image-based analysis, we gauge the predictive capacity of machine learning models trained on visual data. 
        Table \ref{tab:r2-img-id} outlines the R2 scores after 50 epochs, providing a comprehensive overview of model performance. 
        Notably, Xception stands out with a noteworthy R2 score of 0.385, signifying its relatively effective ability to predict prices based on visual attributes.
        
            \begin{table}[htbp]
              \centering
              \caption{R2 Scores by Image machine learning model after 50 epochs}
              \begin{tabular}{cc}
                \toprule
                \textbf{Model} & \textbf{R2 Score} \\
                \midrule
                VGG & 0.320 \\
                EfficientNet & 0.364 \\
                Xception & \textbf{0.385} \\
                DenseNet & 0.341 \\
                ResNet & 0.362 \\
                Inception & 0.351 \\
                \bottomrule
              \end{tabular}
              \label{tab:r2-img-id}
            \end{table}
    
        However, a critical insight emerges when examining solely unseen collections.
        Table \ref{tab:r2-img-ood} showcases the R2 scores in the \textbf{out-of-distribution} context, revealing a consistent negative trend across all models.
        The negative R2 scores indicate a challenge in the models' capacity to generalize effectively to novel and unfamiliar collections.
            \begin{table}[htbp]
                  \centering
                  \caption{R2 Scores by Image machine learning model after 50 epochs with only unseen collections in  the testset}
                  \begin{tabular}{cc}
                    \toprule
                    \textbf{Model} & \textbf{R2 Score} \\
                    \midrule
                    VGG & -0.306639 \\
                    EfficientNet & -0.253686 \\
                    Xception & -0.306964 \\
                    DenseNet & -0.350500 \\
                    ResNet & -0.154064 \\
                    Inception & -0.250115 \\
                    \bottomrule
                  \end{tabular}
              
              \label{tab:r2-img-ood}
            \end{table}

           In essence, the image-centric analysis conveys a nuanced tale: while select models exhibit potential in price prediction within familiar contexts, their efficacy diminishes when confronted with uncharted territory.
           This intricacy underscores the intricate nature of gleaning universally applicable pricing insights from NFT images, emphasizing the ongoing need for exploration and refinement in this domain.

    \subsection{Off-chain Data}

Results in \cref{tab:off-chain-correlations} unveil correlation coefficients, specifically Pearson and Spearman coefficients, that shed light on the connection between diverse collection metadata and the trading volume of collections.

\begin{table}[htp]
  \centering
  \caption{Correlation coefficients of different collection metadata relative to collection trading volume}
  \begin{tabular}{ccc}
    \toprule
    \textbf{Feature} & \textbf{Pearson (p-value)} & \textbf{Spearman (p-value)}  \\
    \midrule
    Twitter Followers & 0.161 ($P<.001$)  & 0.374 ($P<.001$) \\
    Instagram Followers & 0.018 ($P=0.058$)& 0.082 ($P=0.013$) \\
    Age in Days & 0.165 ($P<.001$)  & 0.092 ($P=0.005$) \\
    Creator Fee & -0.119 ($P<.001$) & -0.089 ($P=0.007$)  \\ 
    Website Visits & 0.315 ($P<.001$) & 0.285 ($P<.001$) \\
    \bottomrule
  \end{tabular}
  
  \label{tab:off-chain-correlations}
\end{table}

These coefficients act as measures of the closeness of the relationship between variables, distinguishing between linear (Pearson) and monotonic (Spearman) relationships. 
Their values range from -1 to 1, where -1 implies a strong negative relationship, 1 signifies a strong positive relationship, and 0 denotes a negligible relationship.
The p-value represents the probability that the observed correlation between variables occurred by chance rather than being a meaningful relationship, where values below 0.05 are seen as statistically significant.

When examining the Twitter Follower metric, both Pearson and Spearman coefficients suggest a modest positive connection with collection trading volume. 
This suggests that collections with a higher number of Twitter followers tend to experience greater overall trading volume.
The statistically significant p-values further reinforce this observation, highlighting the reliability of this relationship.

In terms of Instagram Follower, the situation is a bit nuanced.
Both Pearson and Spearman coefficients show a much weaker relationship, with p-values hovering slightly above statistical significance.
We hypothesize that Twitter has a higher impact than Instagram because Twitter as a medium enables better discussion and the forming of communities around collections, than the image posting based Instagram.
We also want to point out that the dataset for Instagram accounts is a lot smaller than for Twitter accounts as less than half of the collections in our dataset have a connected Instagram account, while for Twitter it is more than 90\%.

Turning our attention to the Age in Days of collections, both Pearson and Spearman coefficients present meaningful connections with trading volume.
These coefficients indicate that as the age of a collection increases, so does its trading volume.
This indicates that our analyzed collections still hold value and get actively traded over time.

In the context of the Creator Fee, a different dynamic emerges.
The negative values of both Pearson (-0.119) and Spearman (-0.089) coefficients suggest that collections with higher creator fees tend to have lower trading volumes.
These relationships are statistically significant, as indicated by p-values below 0.001 and 0.007, respectively.

Lastly, the Website Visits metric displays a robust positive connection with trading volume.
Both Pearson and Spearman coefficients emphasize that collections with higher website visit counts tend to experience more active trading.
The low p-values below 0.001 highlight the statistical significance of these findings.
This suggests that a well-designed and informative website can serve as a gateway to increased engagement and transactions.
For creators and platforms, investing in user-friendly websites could potentially amplify trading volumes.
Please note that compared to the other results, website visits were only collected over a timestpan of 3 months.

In practice, these insights provide actionable guidance for NFT stakeholders. 
By strategically leveraging the power of social media, focusing on longevity, optimizing pricing structures, and enhancing online platforms, creators and investors can enhance trading dynamics.
Additionally, the findings underscore the importance of holistic strategies that consider multiple metadata attributes in tandem to maximize trading volume and overall market impact.



\begin{figure}[ht]
        \centering
        \includegraphics[width=\columnwidth]{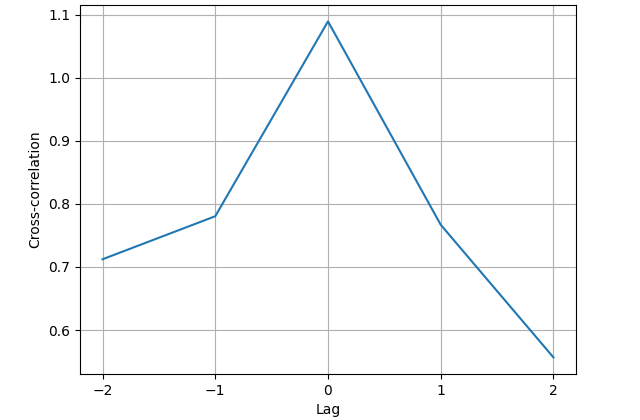}
        \caption{Cross-correlation lags of trends keyword search and monthly trading volume are shown. The vertical axis captures the normalized cross-correlation values, while the horizontal axis denotes the time lags. The highest correlation value emerges at lag 0, signifying data alignment without time-shift.}
        \label{fig:Crosscorrelation-lags}
    \end{figure}

Lastly, we illustrate in \cref{fig:Crosscorrelation-lags} how shifts in keyword popularity align with trading activity over time.

Central to this visualization is cross-correlation, which measures how two signals align over time. 
The vertical axis displays normalized cross-correlation values, while the horizontal axis represents time lags.
It is worth noting that the highest correlation occurs at a 0-lag, indicating an immediate alignment between keyword search and trading volume.
Moreover, as the time lag moves beyond 0, the cross-correlation weakens, highlighting the responsiveness of trading volume to keyword trends.

The zero lag for monthly data suggests that changes in keyword search trends and trading volume occur concurrently with a delay of less than a month.
In practice, this suggests that when there is an increase or decrease in the popularity of certain keywords relevant to the Non-Fungible Token (NFT) market, there is an immediate corresponding effect on trading volume.
This immediate response underscores the rapid and direct relationship between shifts in keyword interest and trading activity within the NFT ecosystem.

\section{Conclusion}

In summary, our exhaustive analysis provides a nuanced view into the complex interplay of factors affecting the pricing and trading volume of NFTs. Utilizing a multifaceted approach that incorporates image analytics, textual metadata, and a host of off-chain elements, we have unearthed valuable insights that shape our understanding of NFT market behavior.

Interestingly, our results challenge some conventional wisdom: while the attributes embedded in the image and text descriptions of NFTs seem to have limited impact on pricing, off-chain factors, paradoxically, exhibit considerable influence on trading volumes. Metrics such as social media reach and web traffic emerge as powerful determinants, showing a statistically significant correlation with trading activities. The role of the creator's fee—a factor intrinsic to the NFT minting process—is also revealed to be a critical variable influencing trade volumes.

These findings unravel the intricacies of the NFT market, elucidating the delicate balance of influences that shape trading behavior and value assignment. As such, our study not only enriches the academic understanding of NFT market mechanisms but also furnishes practical insights for creators, investors, and platforms aiming to optimize their strategies in this rapidly evolving digital frontier. In an ecosystem where the parameters of value and exchange are still fluid, our research stands as an invaluable cornerstone, facilitating more informed and judicious participation in this dynamic marketplace.
\newpage
\bibliography{references}{}
\bibliographystyle{IEEEtran}

\end{document}